\documentclass[twocolumn,english,aps,prl]{revtex4}
\usepackage[T1]{fontenc}
\usepackage[latin9]{inputenc}
\usepackage{amsmath}
\usepackage{amssymb}
\usepackage{graphicx}
\usepackage{esint}

\makeatletter
\@ifundefined{textcolor}{}
{%
 \definecolor{BLACK}{gray}{0}
 \definecolor{WHITE}{gray}{1}
 \definecolor{RED}{rgb}{1,0,0}
 \definecolor{GREEN}{rgb}{0,1,0}
 \definecolor{BLUE}{rgb}{0,0,1}
 \definecolor{CYAN}{cmyk}{1,0,0,0}
 \definecolor{MAGENTA}{cmyk}{0,1,0,0}
 \definecolor{YELLOW}{cmyk}{0,0,1,0}
}

\usepackage{babel}

\makeatother

\usepackage{babel}
\begin{document}
\global\long\def\oc#1{\hat{c}_{#1}}
 \global\long\def\ocd#1{\hat{c}_{#1}^{\dagger}}
 \global\long\def\tr{\text{Tr}\,}
 \global\long\def\im{\text{Im}\,}
 \global\long\def\re{\text{Re}\,}
 \global\long\def\bra#1{\left\langle #1\right|}
 \global\long\def\ket#1{\left|#1\right\rangle }
 \global\long\def\braket#1#2{\left.\left\langle #1\right|#2\right\rangle }
 \global\long\def\obracket#1#2#3{\left\langle #1\right|#2\left|#3\right\rangle }
 \global\long\def\proj#1#2{\left.\left.\left|#1\right\rangle \right\langle #2\right|}

\title{Slow Dynamics in a Two-Dimensional Anderson--Hubbard Model}

\author{Yevgeny Bar Lev}

\author{David R. Reichman}

\affiliation{Department of Chemistry, Columbia University, 3000 Broadway, New
York, New York 10027, USA }
\begin{abstract}
We study the real-time dynamics of a two-dimensional Anderson--Hubbard
model using nonequilibrium self-consistent perturbation theory within
the second-Born approximation. When compared with exact diagonalization
performed on small clusters, we demonstrate that for strong disorder
this technique approaches the exact result on all available timescales,
while for intermediate disorder, in the vicinity of the many-body
localization transition, it produces quantitatively accurate results
up to nontrivial times. Our method allows for the treatment of system
sizes inaccessible by any numerically exact method and for the complete
elimination of finite size effects for the times considered. We show
that for a sufficiently strong disorder the system becomes nonergodic,
while for intermediate disorder strengths and for all accessible time
scales transport in the system is strictly subdiffusive. We argue
that these results are incompatible with a simple percolation picture,
but are consistent with the heuristic random resistor network model
where subdiffusion may be observed for long times until a crossover
to diffusion occurs. The prediction of slow finite-time dynamics in
a two-dimensional interacting and disordered system can be directly
verified in future cold atoms experiments.
\end{abstract}
\maketitle
Ergodicity plays a central role in the statistical mechanics of closed
systems . While it is usually difficult to rigorously prove that a
given system is ergodic, our everyday experience strongly suggests
that generic interacting systems with many-degrees of freedom \emph{are}
ergodic. The underlying assumption is that inelastic collisions between
particles allow for redistribution of energy and relaxation to equilibrium.
This is precisely the reason why it was commonly believed that Anderson
localization \cite{Anderson1958b}, a truly nonergodic phenomenon,
will be destroyed by the addition of a the smallest local inelastic
interactions \cite{Fleishman1978}. It therefore came as a surprise
when two groups argued that a truly nonergodic phase, later dubbed
the many-body localized phase, generically exists at a \emph{finite}
energy density and \emph{finite} interaction strengths \cite{Gornyi2005,Basko2006a}.
By tuning the parameters of the system a dynamical many-body localization
transition between ergodic and nonergodic phases occurs. This transition
was first observed numerically \cite{oganesyan_localization_2007,Pal2010a},
and more recently in cold-atoms experiment \cite{Schreiber2015a}. 

While the existence of a stable nonergodic phase at finite energy
density is the hallmark of many-body localization, transport, or the
absence of thereof, in this phase is not very interesting. On the
other hand, it was recently demonstrated that transport within the
ergodic phase, previously believed to be diffusive \cite{Gornyi2005,Basko2006a},
exhibits subdiffusion for nontrivial times in \emph{one-dimensional}
systems \cite{BarLev2014,Lev2014,Agarwal2014}. A heuristic theoretical
explanation of the subdiffusive behavior suggests that transport in
the system is similar to the random resistor problem \cite{Alexander1981,Alexander1981a},
where rare and large insulating regions have a dominant effect in
one-dimension \cite{Agarwal2014}. Subdiffusion also naturally arises
in the phenomenological renormalization group (RG) treatments of the
same problem, where close to the MBL transition blocking (or thermal)
regions appear on every length scale (RG step) \cite{Vosk2014,Potter2015}.
The fractal structure of the blocking (or thermal) regions has led
to the proposal that the many-body localization transition is a sort
of a percolation transition \cite{Potter2015,Chandran2015,Chandran2015a}.
Similar suggestions were previously made for the metal--insulator
transition in two dimensional systems at zero temperature \cite{Meir2001}.

One of the predictions within the percolation picture is that the
observed subdiffusive behavior in the ergodic phase in one-dimensional
systems will be absent in \emph{higher dimensions}, where blocking
regions can be avoided \cite{Potter2015,Chandran2015,Chandran2015a,Gopalakrishnan2015}.
This prediction is asymptotic in its nature, and any generalization
to finite times will depend on the chosen model. One of the main problems
in studying subdiffusion is that most numerically exact methods struggle
within the ergodic phase. This is in contrast to the case of the nonergodic
MBL phase, where many numerically exact methods are available and
efficient (see \cite{Altman2014,Nandkishore2014} and references within),
. One exception is the study of dc conductivity at low temperatures
using determinantal quantum Monte-Carlo (DQMC), where it was possible
to circumvent the need for analytic continuation and compute dc conductivity
directly in real time. This allowed for the study of the metal--insulator
transition as a function of the interaction strength for a two-dimensional
model \cite{Denteneer1999,Enjalran2001,Chakraborty2007}. It is not
clear, however, if this approach carries over to high temperatures,
of interest to MBL. Moreover, it provides only asymptotic (dc) information.

In this work, we examine the dynamics of a two-dimensional Anderson--Hubbard
model using nonequilibrium perturbation theory within the second-Born
approximation. While our approach is approximate, we show that it
quantitatively reproduces exact results in small systems for a wide
range of parameters and appears to become exact for the timescales
of investigation in the limit of strong disorder. This approach is
similar to the diagrammatic approach of Ref.~\cite{Basko2006a},
which first established the many-body localization transition. We
relax several of the approximations of Ref.~\cite{Basko2006a} and
compute in detail the dynamics of the system from an appropriately
chosen initial condition. 

\begin{figure}
\centering{}\includegraphics[bb=0bp 0bp 243bp 151bp]{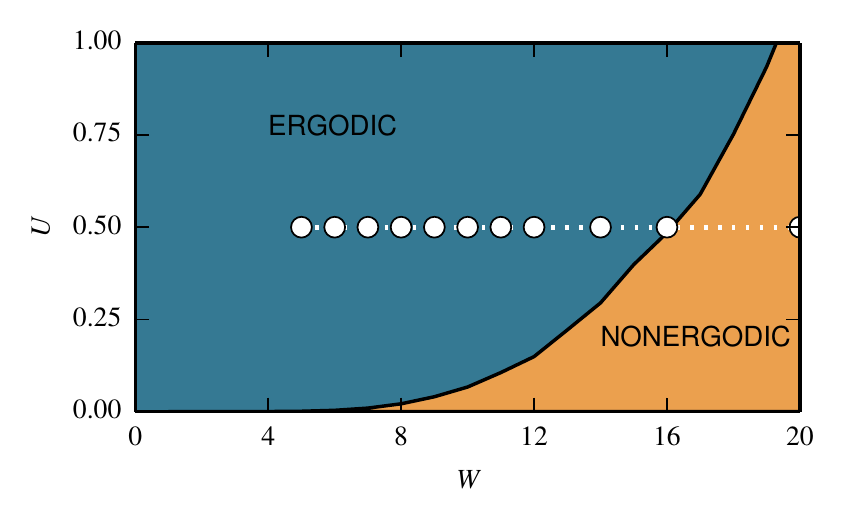}\caption{\label{fig:params}(color online) Dynamical phase boundaries of the
two-dimensional disordered Hubbard model at infinite temperature as
a function of $W$ and $U$. The phase boundary was obtained using
Eq. (93) of Ref.~\cite{Ros2014} and demarcates ergodic and nonergodic
regions. The white circles correspond to the parameters used in this
work $W=5-20$ and $U=0.5$. }
\end{figure}

We investigate a two-dimensional Anderson--Hubbard model,

\begin{eqnarray}
H & = & -t\sum_{\left\langle \mathbf{rr}'\right\rangle ,\sigma}\hat{c}_{\mathbf{r}\sigma}^{\dagger}\hat{c}_{\mathbf{r'}\sigma}+U\sum_{\mathbf{r}}\hat{n}_{\mathbf{r}\uparrow}\hat{n}_{\mathbf{r}\downarrow}+\sum_{\mathbf{r},\sigma}h_{\mathbf{r}\sigma}\hat{n}_{\mathbf{r}\sigma},\label{eq:model}
\end{eqnarray}
where $\hat{c}_{\mathbf{r}\sigma}^{\dagger}$ creates a fermion of
spin $\sigma=\pm1/2$ at site $\mathbf{r}$, $\hat{n}_{\mathbf{r}\sigma}$
is the density operator, $t$ is the hopping matrix element (we set
$t=1$ throughout), $U$ is the interaction strength and $h_{i\sigma}$
are random fields independently distributed on the interval $h_{i\sigma}\in\left[-W,W\right]$.
To eliminate all symmetries of this Hamiltonian we subject the fermionic
species \emph{to different} disorder fields. Since the model has a
finite number of states per site the energy density is bounded, which
allows for meaningful consideration of the infinite temperature limit.
Note that while for systems with unbounded energy density the infinite
temperature limit normally coincides with the classical limit, this
is not necessarily the case for systems with bounded energy density.
For example in the \emph{noninteracting} Anderson model in one and
two dimensions, all the eigenstates are localized and therefore localization
persists at \emph{any} temperature. In this limit the many-body localization
transition occurs as a function of the parameters of the system, and
not the temperature \cite{oganesyan_localization_2007}. In what follows
we will consider only the infinite temperature limit, and use the
strength of the disorder as a control parameter of the many-body localization
transition (we fix the interaction to be $U=0.5$). The theoretical
boundary of the MBL transition at half-filling and infinite temperature
may be estimated as (Eq.~(93) of Ref.~\cite{Ros2014}),
\begin{equation}
U_{c}=0.174\frac{\nu}{K\ln K}\qquad K=4\Delta\nu\xi^{2},\label{eq:trans-line}
\end{equation}
where $\nu,$ $\Delta$ and $\xi$ are the one-particle density of
states, bandwidth and localization length, correspondingly. Taking
the appropriate values for the one-particle problem, the boundary
line of the MBL transition can be readily calculated (see Fig.~\ref{fig:params}).
Since we are interested in studying the dynamics across the transition,
we vary the disorder strength in the interval $5\leq W\leq20$ (yielding\emph{
}a\emph{ non-interacting} localization length $\xi<5.5$). 

To study the dynamics we use the equations of motion for the one-particle
nonequilibrium correlators, 
\begin{eqnarray}
G_{ij}^{>}\left(t;t'\right) & = & -i\tr\left\{ \hat{\rho}_{0}\hat{c}_{i}\left(t\right)\hat{c}_{j}^{\dagger}\left(t'\right)\right\} ,\label{eq:gtr_less_G}\\
G_{ij}^{<}\left(t;t'\right) & = & i\tr\left\{ \hat{\rho}_{0}\hat{c}_{j}^{\dagger}\left(t'\right)\hat{c}_{i}\left(t\right)\right\} ,\nonumber 
\end{eqnarray}
where $\hat{\rho}_{0}$ is the initial density matrix. For an \emph{uncorrelated}
initial density matrix, the Green's functions obey the Kadanoff--Baym
equations of motion \cite{Kadanoff1994}, 
\begin{eqnarray}
i\partial_{t}G^{\gtrless}\left(t,t'\right) & = & \left(\hat{h}_{0}+\Sigma^{HF}\left(t\right)\right)G^{\gtrless}\left(t,t'\right)\nonumber \\
 & + & \int_{0}^{t}\Sigma^{R}\left(t,t_{2}\right)G^{\gtrless}\left(t_{2},t'\right)\mathrm{d}t_{2}\nonumber \\
 & + & \int_{0}^{t'}\Sigma^{\gtrless}\left(t,t_{2}\right)G^{A}\left(t_{2},t'\right)\mathrm{d}t_{2},\label{eq:KB_eq}
\end{eqnarray}
where spatial indices and summations are suppressed for clarity; $\hat{h}_{0}$
is the one particle Hamiltonian; $\Sigma^{HF}\left(t\right)$, $\Sigma^{\gtrless}\left(t\right)$
are the Hartree-Fock greater and lesser self-energies of the problem
respectively; and the superscripts 'R' and 'A' represent retarded
and advanced Green's functions and self-energies, which are defined
as 
\begin{eqnarray}
\Sigma^{R}\left(t,t_{2}\right) & = & \theta\left(t-t_{2}\right)\left(\Sigma^{>}\left(t,t_{2}\right)-\Sigma^{<}\left(t,t_{2}\right)\right)\\
G^{A}\left(t_{2},t'\right) & = & -\theta\left(t'-t_{2}\right)\left(G^{>}\left(t_{2},t'\right)-G^{<}\left(t_{2},t'\right)\right).\nonumber 
\end{eqnarray}
We note in passing that this equation is \emph{exact} if the exact
self energies are employed. Since normally this is not possible, approximate
forms for the self energies are used. Here we use the second-Born
approximation for the self-energy,
\begin{eqnarray}
\Sigma_{ij}^{HF}\left(t\right) & = & -iU\delta_{ij}G_{ii}^{<}\left(t;t\right)\nonumber \\
\Sigma_{ij}^{>}\left(t,t'\right) & = & U^{2}G_{ji}^{<}\left(t',t\right)G_{ij}^{>}\left(t,t'\right)G_{ij}^{>}\left(t,t'\right)\label{eq:self-energies}
\end{eqnarray}
which is a self-consistent conserving approximation, specifically
it conserves the total energy and number of particles and amounts
to a resummation of an infinite class of diagrams. This approximation
was originally considered in the pioneering work of Ref.~\cite{Basko2006a},
but due to the complexity of (\ref{eq:KB_eq}), it was reduced to
a corresponding quantum Boltzmann equation. It is numerically feasible
to solve (\ref{eq:KB_eq}), within full second-Born approximation,
however this is known to generate spurious relaxation \cite{VonFriesen2009}.
This problem can be avoided via the the introduction of the generalized
Kadanoff-Baym anzatz for the greater (lesser) correlator, 
\begin{equation}
G^{\lessgtr}\left(t,t'\right)=i\left[G^{R}\left(t,t'\right)G^{\lessgtr}\left(t',t'\right)-G^{\lessgtr}\left(t,t\right)G^{A}\left(t,t'\right)\right],\label{eq:GKBA}
\end{equation}
and by approximating the retarded and advanced Green's functions with
their Hartree-Fock (HF) values \cite{Spicka2005,Spicka2005a,Latini2013}.
Substituting Eqs.~(\ref{eq:GKBA}) and (\ref{eq:self-energies})
into (\ref{eq:KB_eq}) one obtains the Quantum Master Equation (QME)
for the one-particle density matrix.Unlike the Boltzmann equation
used in Ref.~\cite{Basko2006a}, this approach considers the full
one-particle density matrix and not just its diagonal values, and
therefore includes additional quantum information. We refer the reader
to Refs.~\cite{Spicka2005,Spicka2005a} for additional discussion.

In the infinite temperature limit, the equilibrium density matrix
$\rho_{0}$ in (\ref{eq:gtr_less_G}) is proportional to the unity
matrix, and it is tempting to use this maximally mixed state as an
initial condition. However as alluded to in Ref.~\cite{Basko2006a}
and we explained in our previous work \cite{BarLev2014}, this choice
would lead to erroneous results, in particular in the limit of infinite
disorder. To alleviate this issue we use pure and uncorrelated initial
conditions. Specifically, we fix the number of fermions for each species
(we use half-filing throughout this work) and distribute all fermions
randomly on a square lattice. For these initial conditions it is straightforward
to show that our approach becomes \emph{exact} in the zero hopping
limit. Moreover, as we demonstrate below, even for finite hopping
and large disorder our method quantitatively approaches the exact
behavior exhibited by finite sized systems studied with exact diagonalization.
The infinite temperature limit is achieved by averaging the observables
over random realizations of the initial conditions. To study the dynamics
of the system we calculate the density-density correlation function
at infinite temperature,
\begin{equation}
C\left(\mathbf{r},\mathbf{r}';t\right)=\frac{1}{2Z}\sum_{\alpha\sigma}\left\langle \alpha\left|\left(\hat{n}_{\mathbf{r}\sigma}\left(t\right)-\frac{1}{2}\right)\left(\hat{n}_{\mathbf{r'\sigma}}\left(0\right)-\frac{1}{2}\right)\right|\alpha\right\rangle ,\label{eq:den-den-corr}
\end{equation}
where $\ket{\alpha}$ states are the random initial states described
above and $Z$ is the total number of many-body states. Normally this
quantity requires \emph{two}-particle correlators, however our initial
conditions have the useful property, $\left\langle \hat{n}_{\mathbf{r}\sigma}\left(t\right)\hat{n}_{\mathbf{r}'\sigma}\left(0\right)\right\rangle =\left\langle \hat{n}_{\mathbf{r}\sigma}\left(t\right)\right\rangle n_{\mathbf{r}'\sigma}\left(0\right)$,
where $n_{\mathbf{r}'\sigma}\left(0\right)=0,1$ is the occupation
number of the initial state and,
\begin{equation}
\left\langle \hat{n}_{\mathbf{r}\sigma}\left(t\right)\right\rangle =-iG_{\sigma}^{<}\left(\mathbf{r}t,\mathbf{r}t\right).
\end{equation}
To study the relaxation of the system we calculate the autocorrelation
of the density and the mean-square displacement, 
\begin{equation}
\rho\left(t\right)=\frac{1}{V}\sum_{\mathbf{r}}\overline{C\left(\mathbf{r},\mathbf{r};t\right)},\qquad r^{2}\left(t\right)=\sum_{\mathbf{r}}r^{2}\overline{C\left(\mathbf{r},0;t\right)}.\label{eq:measure}
\end{equation}
For a diffusive system we have, $\rho\left(t\right)\sim t^{-d/2}$,
and $r^{2}\left(t\right)\sim t$, while for localized systems both
$\rho\left(t\right)$ and $r^{2}\left(t\right)$ are expected to saturate
at infinite times to finite quantities. Similar quantities based on
the one-time density $\left[n_{\mathbf{r}}\left(t\right)\right]$
were studied for the clean systems out-of-equilibrium \cite{Fabricius1998,Langer2009,Karrasch2014},
however, such quantities cannot be directly used in equilibrium where
any one-time operator is independent of time. 

\begin{figure}
\begin{centering}
\includegraphics{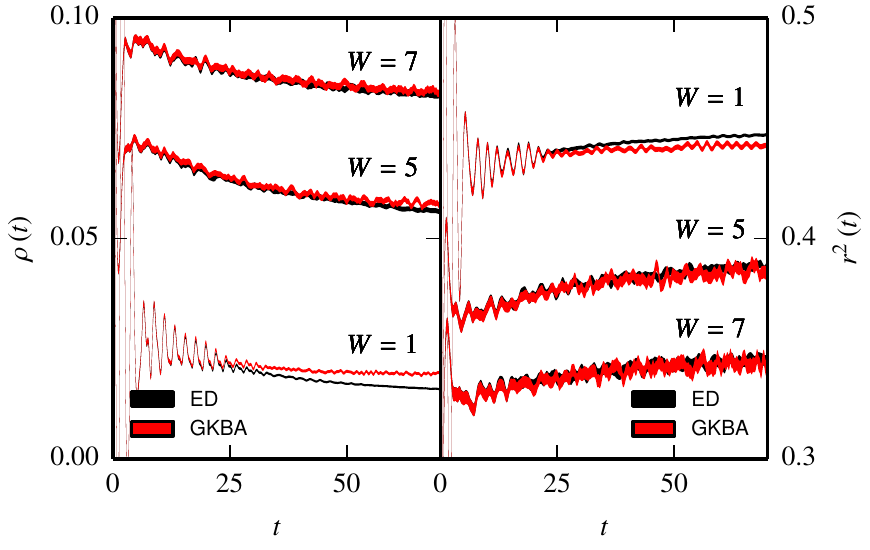} 
\par\end{centering}

\caption{\label{fig:exact-2B}(color online) Comparison between exact diagonalization
(black) and nonequilibrium self-consistent perturbation theory (red/gray)
for several disorder strengths ($W=1,5$ and $7$) performed on a
small Hubbard cluster of dimensions $3\times3$. The left panel shows
the density-density correlation function as a function of time, and
the right shows the average mean-square displacement, computed using
Eq.~(\ref{eq:measure}). An average over $1000$ disorder realizations
was performed and shaded areas designate uncertainty bounds.}
\end{figure}

The quantum master equation (QME) is solved numerically for the one-particle
correlators as a function of time within the self-consistent second-Born
approximation. For this purpose we utilize the parallel numerical
scheme developed in Ref.~\cite{Stan2009}. After obtaining the one-particle
correlators, we extract the density-density correlation function (\ref{eq:den-den-corr})
and average it over both the initial conditions and disorder realizations.
Depending on the strength of the disorder, we use between 35 realizations
for the weakest disorder strengths and 400 realization for the strongest
disorder strengths. The perturbation theory reproduces the exact (and
trivial) result for zero hopping. To evaluate its performance for
nonzero hopping we compare the perturbative calculation to the exact
diagonalization (ED) solution of a small Hubbard cluster of size $3\times3$.
As seen from Fig.~\ref{fig:exact-2B} a remarkable correspondence
between the QME and the exact solutions is seen for $U=0.5$ and times
$t\lesssim40$. This correspondence does \emph{not} exist at the HF
level (not shown). While this comparison has little bearing on the
dynamics of MBL in the thermodynamic limit we will assume that this
correspondence does \emph{not} become worse for larger systems sizes.

The nature of the transport is assessed by examining the finite time
dynamical exponent,
\begin{equation}
\alpha\left(t\right)\equiv\frac{\mathrm{d}\ln r^{2}\left(t\right)}{\mathrm{d}\ln t},\label{eq:dynam_exp}
\end{equation}
which has values $\alpha\left(t\to\infty\right)=2$ for ballistic
transport and $\alpha\left(t\to\infty\right)=1$ for diffusive transport.
This exponent is directly related to the dynamical exponent, $z=2/\alpha$,
which is commonly used in the context of phase transitions and defines
the scaling between time and space, $t\propto L^{z}$. For finite
systems, asymptotic time dynamics will be dictated by finite size
effects such as reflections from the boundaries and the asymptotic
value of $\alpha\left(t\to\infty\right)$ will have little meaning.
Nevertheless, due to the locality of the interaction it takes a finite
time, $t_{*},$ for a local perturbation to reach the boundary of
the system \cite{Lieb1972}. Therefore for times $t<t_{*}$ finite-size
effects can be effectively eliminated. In this work we have chosen
$t_{*}=50$, which is sufficient to remove finite size effects for
the chosen parameters and sizes studied. Such calculations required
about $300,000$ computer hours For the lowest disorder strength we
investigated systems of size $24\times24$ were required while for
the slowest dynamics (strongest disorder strength) systems of the
size $16\times16$ were sufficient. The reader is referred to our
previous works for a more detailed explanation of this procedure \cite{BarLev2014,Lev2014}.
We note in passing that the required system sizes lie far above the
accessible sizes of any \emph{exact} numerical method. Our method
is \emph{approximate,} however Fig.~\ref{fig:exact-2B} suggests
that for the studied disorder strengths and considered times it is
essentially \emph{quantitatively exact}. While the chosen time scale
of $t_{*}=50$ is definitely not asymptotic, as we will show below
it allows access to nontrivial dynamics across the MBL transition.
Moreover this is the relevant time-scale for current experiments in
cold-atoms, where finite size effects from the harmonic trap, particle
loss and decoherence set-in on longer time scales \cite{Lucioni2013,Schreiber2015a}.

\begin{figure}
\begin{centering}
\includegraphics{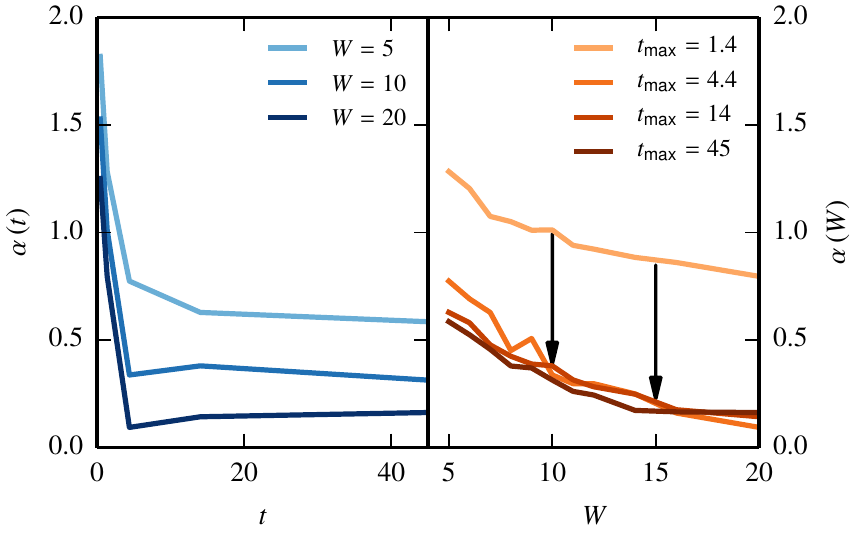}
\par\end{centering}

\caption{(color online) \label{fig:Dynam_Exp}Dynamical exponent across the
many-body localization transition. On the left panel the dynamical
exponent is determined as a function of time for different disorder
strengths. Darker shades designate stronger disorder. The right panel
shows the dynamical exponent as a function of the disorder strength.
Darker shades represent longer times, and the arrows illustrate the
convergence of the dynamical exponent in time.}
\end{figure}
To evaluate the dynamical exponent as a function of time {[}see (\ref{eq:dynam_exp}){]}
, we divided the data into equal time intervals (on a logarithmic
scale) and performed piecewise linear fitting. This is a well defined
procedure, which produces a \emph{time-dependent} dynamical exponent,
and is \emph{different} from numerical extrapolation performed using
power-law fitting. Both procedures however have little bearing on
asymptotic transport. The result is demonstrated for several disorder
values in the left panel of Fig.~\ref{fig:Dynam_Exp}. For localized
systems the value of the dynamical exponent as a function of time,
$\alpha\left(t\right)$, should decay to zero, while for diffusive
systems it should asymptotically saturate to $\alpha\left(t\to\infty\right)=1$.
As clearly seen from Fig.~\ref{fig:Dynam_Exp} neither occurs for
the studied system, instead the dynamical exponent saturates to a
\emph{finite} value on the studied timescale. We repeat this analysis
for various disorder strengths and show that the dynamical exponent
converges to a finite plateau value for all cases under investigation
in the right panel of Fig.~\ref{fig:Dynam_Exp}. This figure summarizes
the main result of our work: for nontrivial times the dynamical exponent
is \emph{finite} across the many-body localization transition and
deep in the ergodic phase. This means that for experimentally relevant
times and a broad range of parameters the system exhibits subdiffusive
transport. While the transition to diffusion can and may occur at
later times (unaccessible to us) , as happens in the case of classical
supercooled liquids \cite{Binder2005}, we do not observe this crossover
in our study even for the lowest disorder strength $\left[W=5\right]$.
Studying even lower disorder strengths is problematic due to the exponential
dependence of the \emph{non-interacting} localization length on the
disorder strength in two-dimensional systems, $\xi\propto\exp\left[A/W^{2}\right]$
\cite{MacKinnon1981}. From our data it is difficult to pinpoint the
location of the MBL transition since $\alpha\left(t\right)$ is never
exactly zero. Nevertheless for $W=15-20$, the values of the dynamical
exponent we obtain are zero within the error bounds, which is consistent
with the theoretical estimate presented in Fig.~\ref{fig:params}.
We emphasize that our results do \emph{not} contradict the expectation
that the MBL dynamics within the ergodic phase of one-dimensional
systems is \emph{qualitatively} distinct from that in higher dimensions,
however the duration of apparent subdiffusion we observe is surprising. 

Note that our results are rather different from dynamics close to
the noninteracting Anderson transition \cite{Ohtsuki1997}. In this
case, close to the transition, the system exhibits apparent subdiffusion
with a \emph{constant} dynamical exponent $\alpha=2/d$, for lengths
of the order of the correlation (localization) lengths and then crosses-over
to either diffusion $\left(\alpha=2\right)$ or localization $\left(\alpha=0\right)$.
This hints that the MBL transition is inherently different from the
Anderson transition. It is also interesting to understand how our
results compare to the heuristic percolation picture \cite{Vosk2014,Potter2015,Chandran2015,Chandran2015a}.
For classical percolation, the dynamics of the system at the percolation
threshold is subdiffusive on all time scales with the universal dynamical
exponent {[}$\alpha\approx0.69$, for $d=2${]}. Above the percolation
threshold $\alpha\left(t\right)$ monotonically increases until it
reaches one (normal diffusion), and below the percolation threshold
$\alpha\left(t\right)$ monotonically decreases to zero \cite{Saxton1994}.
Our results therefore can only correspond to the \emph{nonergodic}
phase \emph{below} the percolation threshold. This renders a literal
mapping of the MBL transition to a classical percolation transition
as suggested in Refs.~\cite{Chandran2015,Chandran2015a} inconsistent
with the theory of Refs.~\cite{Basko2006a,BarLev2014,Ros2014}, which
places the $W=5$ point deep within the ergodic phase (see Fig.~\ref{fig:params}).
On the other hand the random-resistor network picture \cite{Agarwal2014}
can be consistent with our results. Within this picture for a sparse
distribution of the resistors, an initial monotonic decrease of $\alpha\left(t\right)$
can display long plateaus and an eventual crossover to diffusion \cite{Khripkov2014}.
While we have \emph{not} observed this crossover even for the lowest
studied disorder, we \emph{do} observe long plateaus of $\alpha\left(t\right)$
which can signal a precursor of this behavior.

In summary we have studied the nonequilibrium dynamics of a two-dimensional
Anderson--Hubbard model. For this purpose we have utilized self-consistent
nonequilibrium perturbation theory. Our method allows for the study
of large system sizes and the elimination of finite size effects up
to nontrivial times. While our method is approximate, it appears to
become exact in the limit of strong disorder and gives quantitatively
reliable results even for intermediate disorder strengths when compared
to exact diagonalization studies on small systems. We present evidence
consistent with the existence of the many-body localized phase at
a two-dimensional system at infinite temperature, and put an upper
bound on the location of the many-body localization for a particular
interaction strength. Surprisingly, up to the studied times we find
subdiffusive dynamics for all the studied parameters. We compare our
results to the predictions of the available heuristic theories and
point out that while they can be explained within the random-resistor
network picture they are inconsistent with the literal adoption of
classical percolation theory for the MBL transition. Our results predict
that experiments performed on cold atoms in two-dimensional disordered
systems will observe a finite regime of subdiffusive transport for
times shorter than the typical decoherence time in these experiments.
\begin{acknowledgments}
We would like to thank G. Cohen for many enlightening and helpful
discussions. This work was supported by grant NSF-CHE-1644802.
\end{acknowledgments}

\bibliographystyle{apsrev}
\bibliography{/home/yevgeny/Dropbox/Research/Latex/papers/Bibs/library,local}

\end{document}